\documentclass[sigconf]{acmart}
\AtBeginDocument{%
  \providecommand\BibTeX{{%
    \normalfont B\kern-0.5em{\scshape i\kern-0.25em b}\kern-0.8em\TeX}}}

\copyrightyear{2022} 
\acmYear{2022} 
\setcopyright{acmcopyright}\acmConference[MSR '22]{19th International Conference on Mining Software Repositories}{May 23--24, 2022}{Pittsburgh, PA, USA}
\acmBooktitle{19th International Conference on Mining Software Repositories (MSR '22), May 23--24, 2022, Pittsburgh, PA, USA}
\acmPrice{15.00}
\acmDOI{10.1145/3524842.3528016}
\acmISBN{978-1-4503-9303-4/22/05}

\acmConference[MSR 2022]{MSR '22: Proceedings of the 19th International Conference on Mining Software Repositories}{May 23–24, 2022}{Pittsburgh, PA, USA}
\begin{document}

\title{Dataset: Dependency Networks of Open Source Libraries Available Through CocoaPods, Carthage and Swift PM}

\author{Kristiina Rahkema}
\orcid{0000-0001-7332-2041}
\affiliation{%
  \institution{University of Tartu}
  \city{Tartu}
  \country{Estonia}
  \postcode{51009}
}
\email{kristiina.rahkema@ut.ee}

\author{Dietmar Pfahl}
\affiliation{%
  \institution{University of Tartu}
  \city{Tartu}
  \country{Estonia}
  \postcode{51009}
}
\email{dietmar.pfahl@ut.ee}


\begin{abstract}
Third party libraries are used to integrate existing solutions for common problems and help speed up development. The use of third party libraries, however, can carry risks, for example through vulnerabilities in these libraries. Studying the dependency networks of package managers lets us better understand and mitigate these risks. So far, the dependency networks of the three most important package managers of the Apple ecosystem, CocoaPods, Carthage and Swift PM, have not been studied. We analysed the dependencies for all publicly available open source libraries up to December 2021 and compiled a dataset containing the dependency networks of all three package managers. The dependency networks can be used to analyse how vulnerabilities are propagated through transitive dependencies. In order to ease the tracing of vulnerable libraries we also queried the NVD database and included publicly reported vulnerabilities for these libraries in the dataset. 
\end{abstract}




\maketitle

\section{Introduction}

Third party libraries allow developers to use existing solutions for common tasks and speed up development. For almost every popular programming language there is at least one package manager that can be used to manage these dependencies. 

A recent vulnerability in the popular log4j java logging library affected around four percent of all projects in the Maven repository \cite{log4j}. In 2015, a vulnerability in the popular iOS third party library AFNetworking was found. The vulnerability affected around 1000 iOS applications with millions of users \cite{afnetworking}. Apps can not only be affected when they directly depend on these libraries, but also when their dependencies or dependencies of dependencies depend on these vulnerable library version. 

The dependency network of a package manager contains all libraries distributed through this package manager and their dependency relationships with other libraries. Dependency networks, including their growth and vulnerability, have been studied thoroughly for many package managers such as, among others, npm, RubyGems and Cargo \cite{kikas2017structure, decan2019empirical}. 

Both Decan et al. \cite{decan2019empirical} and Kikas et al. \cite{kikas2017structure} highlighted differences between package managers and how policies and quality of the standard library of a language can affect the dependency network structure. So far the dependency networks for the package managers used in iOS development (CocoaPods, Carthage and Swift Package Manager) have not been studied. 

We start closing this gap by creating a dataset\footnote{\url{https://zenodo.org/record/6376009} \cite{dataset}} containing the dependency networks for libraries provided through CocoaPods, Carthage and Swift Package Manager. The dataset contains information on open source library versions provided through CocoaPods, Carthage and Swift Package Manager, dependency relationships between these versions and information on publicly disclosed vulnerabilities in the NVD database.




\section{Related Work}

There are different ways to construct the dependency network of a package manger. Li et al. \cite{pdgraph} created a project dependency graph for Java Maven projects where each node represents a project and edges between nodes denote the inter-dependency requirements, for example "> 5.0.0" if all versions starting with the major version 5 are allowed. 

Kikas et al. \cite{kikas2017structure} created a dependency graph for JavaScript, Ruby and Rust ecosystems and analysed their evolution. They discussed different approaches for the dependency network construction and highlighted the importance of storing the dependencies between actual versions of libraries as opposed to an aggregated approach where dependencies are connections between library nodes without version information. Dependencies between versions were found by analysing the package manager manifest files. Dependency constraints that allowed the use of multiple library versions were handled by observing which library versions would have been the best match at the time when the current library version was released. 

The biggest dataset containing information on library dependencies is the newest (2020) version of the open dataset provided by libraries.io \cite{librariesio}. The dataset contains data on 32 package managers and three source code repositories. For many pacakge managers the data includes dependency data between all versions of libraries. Data on the corresponding software repositories also contains information on the dependency constraints used, this data is however based on the last snapshot of the project repository. In 2019 Decan et al. \cite{decan2019empirical} studied the evolution of dependency networks for seven different package managers using the 2017 libraries.io dataset. The package managers CocoaPods, Carthage and Swift Package Manager were not included in the study as the dataset only included data on dependencies between software repositories and there was no data on dependencies between library versions. 

The data on dependencies between library versions for CocoaPods, Carthage and SwiftPM is also missing from the newest version of the libraries.io dataset. Data on library versions and dependencies between library versions is however necessary for the analysis of dependency network evolution. 

To fill this gap we compiled a database containing library dependency information on version level for CocoaPods, Carthage and SwiftPM package managers. For a more complete picture we also included data on dependency constraints that can be used similarly to \cite{pdgraph}. Additionally we queried publicly reported vulnerabilities from the NVD database and matched them to the library versions in our dataset. 

\section{Data Collection Methodology}
For data collation it was necessary to identify libraries that are available through each package manager, to collect dependency data for each library and to collect vulnerability data for each library. 

\subsection{Package Managers}
Developers can include third party libraries into projects in multiple ways. Libraries can be either directly downloaded and imported manually or developers can use package managers. Use of a package manager makes it easier to include a new library and foremost it makes it easier to keep the library up to date. The package managers used in Swift development (i.e. for iOS, Mac OS and Watch OS applications) are CocoaPods, Carthage and Swift Package Manager. These package managers are fundamentally different. 

CocoaPods is a package manager with a central database of libraries. If a developer wants to distribute their library through CocoaPods they need to create a Podspec file and add it to the CocoaPods Spec repository. This repository is public and can be accessed by anyone. 

Swift Package Manager (Swift PM) is the official package manager for Swift. It is however not the most popular package manager and compatibility with iOS projects was not added until 2019\cite{swiftpm_ios}. Swift PM does not have a central list of libraries. Any library that includes a Package.swift manifest file can be included through Swift PM by providing its repository address.

Carthage, similarly to Swift PM, is a decentralized package manager. A library can be included by providing its repository address, binary location or path on the local file system. 

\subsection{Identifying Libraries}

Due to the differences in the package managers, we identified libraries for CocoaPods differently than for Carthage and Swift PM. For CocoaPods we cloned the Spec repository\footnote{\url{https://github.com/CocoaPods/Specs}} and extracted all repository URLs from the Podspec files. The extracted list included 79557 repository URLs. Among these URL were incorrect values, such as "./", ".git", "someone@gmail.com". Some URLs were correct URLs, but they were links to private repositories on company domains. 

We discarded all URLs that did not contain github.com or bitbucket.org and 73243 URLs (92\%) remained. We looked trough these values and noticed that some of the URLs contained references to passwords of the form username:password@github.com. We decided to strip these values before the analysis. This leads to us not being able to access some of the repositories. Nevertheless we assumed that it might have been developers intention to not make the library code accessible to everyone, although the password and username combination would be accessible to anyone through the public Spec repository. 

For Carthage and Swift PM we used the libraries.io dataset \cite{librariesio} to get a list of library names. The set of library names is not complete since the dataset was compiled in 2020 and new libraries may have been created after that. We extracted 3880 names for Carthage libraries and 4207 names for Swift PM libraries. 

We ran our analysis on these three sets of libraries. The analysis was successful for 56822 CocoaPods libraries, 2118 Swift PM libraries and 3094 Carthage libraries. We then merged the three databases. The merged database contained 60084 successfully analysed libraries.

We then queried libraries that are referenced as dependencies but that are not analysed. There were a total of 4728 library dependencies of which 1047 were not analysed.  We gathered the names of these libraries and performed a second round of analysis. We refer to this as library snowballing. There may still exist libraries, that are available through one of the three package managers, that are missed by our approach. These libraries however will not have any dependents and would not be essential to dependency network analysis. 

\subsection{Collecting Dependency Data}

Dependency data was collected by parsing the manifest files Package.swift, Podfile and Cartfile and package manager resolution files Package.resolved, Podfile.lock and Cartfile.resolved. We extracted Library names, versions and version constraints from the mainifest files and stored the data as library definitions. We parsed package manager resolution files and extracted library dependencies with names and versions. Package resolution files contain the exact version of each library that the package manager deemed to be the best match at the time when the developer last updated the dependencies. Package resolution files also contain information on all transitive dependencies. Information on both direct and transitive dependencies were stored in the database. 

To match library names extracted from the three package managers it is necessary to "translate" library names. The translation was done by finding information on the library repository URL from the CocoaPods Spec repository. We used the repository username/projectname combination as the library name. 

\subsection{Library Snowballing}
We queried libraries that are referenced as dependencies from other libraries but were not yet analysed. These libraries included open source libraries that had not been analysed yet and closed sourced or local libraries that were not accessible to us. 

For each library name that is a dependency through CocoaPods we were able to extract the repository URL from the Spec repository. These libraries should have been already analysed.
If some of these URLs were discarded previously but the repository is accessible to us the project was included in the analysis. 

For each library name that was not a dependency through CocoaPods, the name was of the form username/projectname. For each name we tried to query the repository. If we were able to access the repository the library was analysed. If we were not able to access the repository the library was ignored. 

The snowballing process added 451 additional libraries to the database.

\subsection{Collecting Vulnerability Data}

For vulnerability data we used the NVD database\footnote{\url{https://nvd.nist.gov}}. For each project that has publicly reported vulnerabilities there is a unique CPE (Common Product Enumeration) value. We downloaded the CPE dictionary\footnote{\url{https://nvd.nist.gov/products/cpe}}. We then went through the dictionary and extracted all repository URLs and their corresponding CPEs. We are analysing open source libraries, therefore we decided to extract repository URLs and ignore all entries that did not include a source reference. We found 5885 CPE values. 

Next, for each library name, we checked if it matched with the entries in the CPE list. We found 51 matching values. For each CPE value we queried the NVD database to find vulnerabilities related to each CPE. We checked each library that was matched to a CPE value to determine if the library was indeed part of the CocoaPods, Carthage and Swift PM ecosystem. We removed two libraries that were included through libraries.io, but were not relevant to the ecosystem. We then matched library versions from the vulnerability data with library versions in our database. We found 159 vulnerabilities in total that affected 41 libraries and 1339 library versions.

\section{Data Collection Tools}
We developed and used the following open source tools for data collection: GraphifyEvolution\footnote{\url{https://github.com/kristiinara/GraphifyEvolution}}, SwiftDependencyChecker\footnote{\url{https://github.com/kristiinara/SwiftDependencyChecker}} and LibraryDependencyAnalysis\footnote{\url{https://github.com/kristiinara/LibraryDependencyAnalysis}}.

\subsection{GraphifyEvolution}
We developed GraphifyEvolution \cite{rahkema2021graphify} for bulk analysis of applications written in Swift. The tool is built in a modular manner and is designed so that it is easy to include analysis results from external analysers. GraphifyEvolution is able to take a list of applications and analyse their evolution based on git commits or tags. For commits the git commit tree is used to determine the evolution between versions. For git tags the evolution between versions is determined by the commit timestamp. Our analysis used the git tag option since tags are normally connected to library versions. All data is entered into a neo4j database.

\subsection{SwiftDependencyChecker}

SwiftDependencyChecker \cite{rahkema2022checker} was developed to detect if an app uses dependencies with publicly reported vulnerabilities. Its intended use is as part of the build process inside Xcode to warn developers about vulnerable dependencies, but it can also be used to detect dependencies declared though CocoaPods, Carthage and Swift Package Manager. We integrated SwiftDependencyChecker into GraphifyEvolution as an external analyser. The external analyser implementation finds dependencies declared for each project version and enters libraries and library definitions into the neo4j as nodes. Relationships are created between project version and its direct and indirect dependencies.

\subsection{LibraryDependencyAnalysis}
We wrote a shell script that first finds libraries that are accessible from CocoaPods, Carthage and Swift PM. It splits the library data into batches and runs GraphfiyEvolution on each batch. The LibraryDependencyAnalysis script can query library data either from libraries.io API\footnote{\url{https://libraries.io/api}}, a postgresql database containing the projects table from the libraries.io data archive\cite{librariesio} or the CocoaPods Spec repository\footnote{\url{https://github.com/CocoaPods/Specs}}. We compiled this database by running the LibraryDependencyAnalysis script with the postgresql option for Carthage and Swift PM and the CocoaPods spec repository option for CocoaPods project.

\section{Dataset Description}


The dataset is provided as a neo4j\footnote{\url{https://neo4j.com/product/neo4j-graph-database/}} database and can be downloaded from the database repository\footnote{\url{https://zenodo.org/record/6376009} \cite{dataset}}. The database dump can be loaded with the following command, where \texttt{--from} points to the database dump file and \texttt{--to} specifies the name of an empty database: 

\begin{verbatim}
    neo4j-admin load --from=<db-dump> --to=<db-name>
\end{verbatim}

The database can be used with either the free community version of the neo4j server or the enterprise version. We also provide the data in json format, which makes it accessible without using neo4j. 

Neo4j is a graph database where data is represented as nodes and relationships. Our database contains the following types of nodes: 

\begin{itemize}
    \item Project: reference to a repository (75323)
    \item App: released version of a project (572131)
    \item Library: resolved dependency with exact version (576144)
    \item LibraryDefinition: dependency definition from manifest file with dependency constraints (19390)
    \item Vulnerability: publicly reported vulnerability from the NVD database (159)
\end{itemize}

Nodes are connected through the following six relationships: 
\begin{itemize}
    \item Project - HAS\_APP -> App (572286)
    \item App - IS -> Library (572122)
    \item App - DEPENDS\_ON -> Library (312454)
    \item App - DEPENDS\_ON\_INDIRECTLY -> Library (267453)
    \item App - DEPENDS\_ON -> LibraryDefinition (383020)
    \item Library - HAS\_VULNERABILITY -> Vulnerability (3761)
\end{itemize}

Detailed descriptions on node and relationship properties is given on the database repository page\footnote{\url{https://github.com/kristiinara/LibraryDependencyAnalysis/\#readme}}. It is possible to generate additional relationships if needed to ease further analyses. For example, adding a DEPENDS\_ON relationship between Library nodes could speed up analysing the length of transitive dependency chains. 

\section{Impact and Research Directions}
The database contains the combined dependency networks for CocoaPods, Carthage and Swift PM. Compared to other dependency networks we also added data on openly reported vulnerabilities for these libraries from the NVD database. 

The database can be used to analyse the dependency network evolution for CocoaPods, Carthage and Swift PM. Dependency network evolution has been studied for many different languages \cite{kikas2017structure,decan2019empirical}, but these three package managers have not been analysed before. This data can be used to analyse the growth and vulnerability of the Swift dependency ecosystem (including libraries written in C, C++, and Objective-C, if so requested).

The additional data on publicly reported vulnerabilities allows the analysis of how vulnerabilities spread through these dependency networks. There is extensive analysis on vulnerabilities in the npm \cite{decan2018impact, zimmermann2019small} dependency network. However these results might not be generalizable to CocoaPods, Carthage and Swift PM. JavaScript is very different from Swift. JavaScript has a very small standard library, whereas the standard library in Swift and the additional system libraries for iOS and Mac OS are very extensive. Swift developers are more wary of including third party libraries in their projects and might be more inclined to update library versions due to limited backwards compatibility of Swift.

\section{Limitations and Challenges}

There were certain limitations on which libraries we were able to analyse and on how exact the vulnerability matching is. 

For SwiftPM and Carthage we relied on the data from the libraries.io dataset. The list of SwiftPM libraries included many libraries where its repository did not include any tags. These libraries could not be included through the package managers by specifying a version and were therefore not analysed. In total 50\% of the Swift PM libraries in the libraries.io dataset were discarded. We checked how many of these libraries were referenced by other libraries and found that 19 referenced libraries had no tags and were therefore not analysed. 

CocoaPods libraries that were not available through GitHub or Bitbucket were discarded. Additionally we ignored any credentials embedded in the repository URL, as we could not be certain that this was intentional. In total, we were not able to analyse 29\% of libraries from the CocoaPods spec repository. If there were any dependencies to these libraries, then they still appear in the database as libraries, but without any additional analysis on their dependencies. 

We only included open source libraries in our analysis. Dependencies to closed sourced libraries are included if they are referenced in the package manager resolution or manifest files, but dependencies of these dependencies are not analysed. A total of 14\% of libraries with dependents are not analysed. 

For some library versions package manager manifest files existed, but there were no package manager resolution files. We did not try to resolve dependencies for these libraries ourselves (13\% of libraries with manifest files). We did however include information on dependencies from the manifest files as library definitions. As a future improvement for these versions the dependency information from the manifest files could be used to calculate the resolved versions at the time when the version was released, as was done in \cite{kikas2017structure}. 

Affected CPEs in the vulnerability database were not listed directly, but as a tree that included operators (such as AND, OR) describing how the product is affected. We ignored the operator values and recorded every CPE value that was specified as vulnerable to be vulnerable. Vulnerability data on these libraries did not seem to include any complex rules on which CPEs are really affected by each vulnerability.

Due to the amount of libraries, finishing the analysis was challenging. In total it took 11 days to run the dependency analysis on all libraries. Fortunately updating this database will take less time, as it is possible to query last commits that were analysed and start the new analysis from that point.

\section{Conclusions}

We compiled a database containing the library dependency network for CocoaPods, Carthage and Swift PM. So far, the dependency networks of these package managers have not been studied yet. There are distinctive differences between Swift and other languages (for example JavaScript) that could have an effect on its third party library dependency network. 

It is necessary to better understand these differences to help developers in protecting their apps against possible vulnerabilities in third party dependencies as effectively as possible. The inclusion of vulnerability data allows us to analyse the real world risks that stem from using outdated dependencies on this platform. 

One of our goals was to make this database as easily accessible as possible. Therefore after the data and neo4j has been downloaded the dataset can be imported with a single command. Additionally all tools used to compile this dataset are open source and available on GitHub. 

The database contains dependency information on 60533 libraries. There are 312454 dependencies between library versions and a total of 159 vulnerabilities. 
It took our script 11 days to analyse all libraries. 



\begin{acks}
Funding of this research came from the Estonian Center of Excellence in ICT research (EXCITE), the European Social Fund via IT Academy program, the Estonia Research Council grant (PRG 1226), the Austrian ministries BMVIT and BMDW, and the Province of Upper Austria under the COMET (Competence Centers for Excellent Technologies) Programme managed by FFG.
\end{acks}

\bibliographystyle{ACM-Reference-Format}
\bibliography{sample-base}


\begin{thebibliography}{12}


\ifx \showCODEN    \undefined \def \showCODEN     #1{\unskip}     \fi
\ifx \showDOI      \undefined \def \showDOI       #1{#1}\fi
\ifx \showISBNx    \undefined \def \showISBNx     #1{\unskip}     \fi
\ifx \showISBNxiii \undefined \def \showISBNxiii  #1{\unskip}     \fi
\ifx \showISSN     \undefined \def \showISSN      #1{\unskip}     \fi
\ifx \showLCCN     \undefined \def \showLCCN      #1{\unskip}     \fi
\ifx \shownote     \undefined \def \shownote      #1{#1}          \fi
\ifx \showarticletitle \undefined \def \showarticletitle #1{#1}   \fi
\ifx \showURL      \undefined \def \showURL       {\relax}        \fi
\providecommand\bibfield[2]{#2}
\providecommand\bibinfo[2]{#2}
\providecommand\natexlab[1]{#1}
\providecommand\showeprint[2][]{arXiv:#2}

\bibitem[Constantin(2015)]%
        {afnetworking}
\bibfield{author}{\bibinfo{person}{Lucian Constantin}.}
  \bibinfo{year}{2015}\natexlab{}.
\newblock \bibinfo{booktitle}{\emph{HTTPS snooping flaw in third-party library
  affected 1,000 iOS apps with millions of users}}.
\newblock
\urldef\tempurl%
\url{https://www.computerworld.com/article/2912402/https-snooping-flaw-in-third-party-library-affected-1000-ios-apps-with-millions-of-users.html}
\showURL{%
Retrieved January 11, 2022 from \tempurl}


\bibitem[Decan et~al\mbox{.}(2018)]%
        {decan2018impact}
\bibfield{author}{\bibinfo{person}{Alexandre Decan}, \bibinfo{person}{Tom
  Mens}, {and} \bibinfo{person}{Eleni Constantinou}.}
  \bibinfo{year}{2018}\natexlab{}.
\newblock \showarticletitle{On the impact of security vulnerabilities in the
  npm package dependency network}. In \bibinfo{booktitle}{\emph{Proceedings of
  the 15th International Conference on Mining Software Repositories}}.
  \bibinfo{pages}{181--191}.
\newblock
\urldef\tempurl%
\url{https://doi.org/10.1145/3196398.3196401}
\showDOI{\tempurl}


\bibitem[Decan et~al\mbox{.}(2019)]%
        {decan2019empirical}
\bibfield{author}{\bibinfo{person}{Alexandre Decan}, \bibinfo{person}{Tom
  Mens}, {and} \bibinfo{person}{Philippe Grosjean}.}
  \bibinfo{year}{2019}\natexlab{}.
\newblock \showarticletitle{An empirical comparison of dependency network
  evolution in seven software packaging ecosystems}.
\newblock \bibinfo{journal}{\emph{Empirical Software Engineering}}
  \bibinfo{volume}{24}, \bibinfo{number}{1} (\bibinfo{year}{2019}),
  \bibinfo{pages}{381--416}.
\newblock
\urldef\tempurl%
\url{https://doi.org/10.1007/s10664-017-9589-y}
\showDOI{\tempurl}


\bibitem[Elliott(2020)]%
        {swiftpm_ios}
\bibfield{author}{\bibinfo{person}{Tom Elliott}.}
  \bibinfo{year}{2020}\natexlab{}.
\newblock \bibinfo{booktitle}{\emph{Swift Package Manager for iOS}}.
\newblock
\urldef\tempurl%
\url{https://www.raywenderlich.com/7242045-swift-package-manager-for-ios}
\showURL{%
Retrieved January 21, 2022 from \tempurl}


\bibitem[Katz(2020)]%
        {librariesio}
\bibfield{author}{\bibinfo{person}{Jeremy Katz}.}
  \bibinfo{year}{2020}\natexlab{}.
\newblock \bibinfo{booktitle}{\emph{{Libraries.io Open Source Repository and
  Dependency Metadata}}}.
\newblock
\urldef\tempurl%
\url{https://doi.org/10.5281/zenodo.3626071}
\showDOI{\tempurl}


\bibitem[Kikas et~al\mbox{.}(2017)]%
        {kikas2017structure}
\bibfield{author}{\bibinfo{person}{Riivo Kikas}, \bibinfo{person}{Georgios
  Gousios}, \bibinfo{person}{Marlon Dumas}, {and} \bibinfo{person}{Dietmar
  Pfahl}.} \bibinfo{year}{2017}\natexlab{}.
\newblock \showarticletitle{Structure and evolution of package dependency
  networks}. In \bibinfo{booktitle}{\emph{2017 IEEE/ACM 14th International
  Conference on Mining Software Repositories (MSR)}}. IEEE,
  \bibinfo{pages}{102--112}.
\newblock
\urldef\tempurl%
\url{https://doi.org/10.1109/MSR.2017.55}
\showDOI{\tempurl}


\bibitem[Li et~al\mbox{.}(2021)]%
        {pdgraph}
\bibfield{author}{\bibinfo{person}{Qiang Li}, \bibinfo{person}{Jinke Song},
  \bibinfo{person}{Dawei Tan}, \bibinfo{person}{Haining Wang}, {and}
  \bibinfo{person}{Jiqiang Liu}.} \bibinfo{year}{2021}\natexlab{}.
\newblock \showarticletitle{PDGraph: A Large-Scale Empirical Study on Project
  Dependency of Security Vulnerabilities}. In \bibinfo{booktitle}{\emph{2021
  51st Annual IEEE/IFIP International Conference on Dependable Systems and
  Networks (DSN)}}. \bibinfo{pages}{161--173}.
\newblock
\urldef\tempurl%
\url{https://doi.org/10.1109/DSN48987.2021.00031}
\showDOI{\tempurl}


\bibitem[Rahkema and Pfahl(2021)]%
        {rahkema2021graphify}
\bibfield{author}{\bibinfo{person}{Kristiina Rahkema} {and}
  \bibinfo{person}{Dietmar Pfahl}.} \bibinfo{year}{2021}\natexlab{}.
\newblock \showarticletitle{GraphifyEvolution-A Modular Approach to Analysing
  Source Code Histories}. In \bibinfo{booktitle}{\emph{Proceedings of the
  IEEE/ACM 8th International Conference on Mobile Software Engineering and
  Systems}}. \bibinfo{pages}{24--27}.
\newblock
\urldef\tempurl%
\url{https://doi.org/10.1109/MobileSoft52590.2021.00009}
\showDOI{\tempurl}


\bibitem[Rahkema and Pfahl(2022a)]%
        {dataset}
\bibfield{author}{\bibinfo{person}{Kristiina Rahkema} {and}
  \bibinfo{person}{Dietmar Pfahl}.} \bibinfo{year}{2022}\natexlab{a}.
\newblock \bibinfo{booktitle}{\emph{{Dependency Networks of Open Source
  Libraries Available Through CocoaPods, Carthage and Swift PM}}}.
\newblock
\urldef\tempurl%
\url{https://doi.org/10.5281/zenodo.6376009}
\showDOI{\tempurl}


\bibitem[Rahkema and Pfahl(2022b)]%
        {rahkema2022checker}
\bibfield{author}{\bibinfo{person}{Kristiina Rahkema} {and}
  \bibinfo{person}{Dietmar Pfahl}.} \bibinfo{year}{2022}\natexlab{b}.
\newblock \showarticletitle{SwiftDependencyChecker: Detecting Vulnerable
  Dependencies Declared Through CocoaPods, Carthage and Swift PM}. In
  \bibinfo{booktitle}{\emph{Proceedings of the IEEE/ACM 9th International
  Conference on Mobile Software Engineering and Systems}}.
\newblock
\urldef\tempurl%
\url{https://doi.org/10.1145/3524613.3527806}
\showDOI{\tempurl}


\bibitem[Wetter and Ringland(2021)]%
        {log4j}
\bibfield{author}{\bibinfo{person}{James Wetter} {and} \bibinfo{person}{Nicky
  Ringland}.} \bibinfo{year}{2021}\natexlab{}.
\newblock \bibinfo{booktitle}{\emph{Understanding the Impact of Apache Log4j
  Vulnerability}}.
\newblock
\urldef\tempurl%
\url{https://security.googleblog.com/2021/12/understanding-impact-of-apache-log4j.html}
\showURL{%
Retrieved January 11, 2022 from \tempurl}


\bibitem[Zimmermann et~al\mbox{.}(2019)]%
        {zimmermann2019small}
\bibfield{author}{\bibinfo{person}{Markus Zimmermann},
  \bibinfo{person}{Cristian-Alexandru Staicu}, \bibinfo{person}{Cam Tenny},
  {and} \bibinfo{person}{Michael Pradel}.} \bibinfo{year}{2019}\natexlab{}.
\newblock \showarticletitle{Small World with High Risks: A Study of Security
  Threats in the npm Ecosystem}. In \bibinfo{booktitle}{\emph{28th USENIX
  Security Symposium (USENIX Security 19)}}. \bibinfo{publisher}{USENIX
  Association}, \bibinfo{address}{Santa Clara, CA}, \bibinfo{pages}{995--1010}.
\newblock
\showISBNx{978-1-939133-06-9}
\urldef\tempurl%
\url{https://www.usenix.org/conference/usenixsecurity19/presentation/zimmerman}
\showURL{%
\tempurl}


\end{thebibliography}


\end{document}